\documentclass[aps,prd,preprint,superscriptaddress,showpacs,nofootinbib]{revtex4-1}

\usepackage{amsmath}
\usepackage{graphicx}
\usepackage{epstopdf}
\usepackage{latexsym}
\usepackage{amssymb}
\usepackage{hyperref}
\usepackage{comment}

\begin{document}


\title{Testing coupled dark energy models with their cosmological background evolution}

\author{Carsten van de Bruck}
\email{c.vandebruck@sheffield.ac.uk}
\affiliation{Consortium for Fundamental Physics, School of Mathematics and Statistics, University of Sheffield, Hounsfield Road, Sheffield S3 7RH, UK} 
\author{Jurgen Mifsud}
\email{jmifsud1@sheffield.ac.uk}
\affiliation{Consortium for Fundamental Physics, School of Mathematics and Statistics, University of Sheffield, Hounsfield Road, Sheffield S3 7RH, UK} 
\author{Jack Morrice}
\email{app12jam@sheffield.ac.uk}
\affiliation{Consortium for Fundamental Physics, School of Mathematics and Statistics, University of Sheffield, Hounsfield Road, Sheffield S3 7RH, UK} 

\date{\today}

\begin{abstract}
We consider a cosmology in which dark matter and a quintessence scalar field responsible for the acceleration of the Universe are allowed to interact. Allowing for both conformal and disformal couplings, we perform a global analysis of the constraints on our model using Hubble parameter measurements, baryon acoustic oscillation distance measurements, and a Supernovae Type Ia data set.
We find that the additional disformal coupling relaxes the conformal coupling constraints. Moreover we show that, at the background level, a disformal interaction within the dark sector is preferred to both $\Lambda$CDM and uncoupled quintessence, hence favouring interacting dark energy.
\end{abstract}


\maketitle



\section{Introduction}
Multiple high precision cosmological observations broaden our understanding of the dynamics of the Universe when confronted with theoretical models. For instance, inferences from observations of Supernovae Type Ia (SNIa) \cite{Perlmutter:1998np,Riess:1998cb,Garnavich:1998th,Schmidt:1998ys,Perlmutter:1997zf}, baryon acoustic oscillations (BAO) \cite{Eisenstein:2005su,Percival:2007yw,Percival:2009xn}, and the cosmic microwave background (CMB) \cite{Spergel:2003cb,Spergel:2006hy,Reichardt:2008ay,Ade:2015xua} are complementary---among other things they indicate that our Universe has recently entered an accelerating epoch. Analysis from data sets of this kind has led cosmologists to formulate a standard model that postulates a dark sector consisting of dark energy and dark matter, contributing to about $69\%$ and $26\%$ of the total energy density in the Universe respectively \cite{Ade:2015xua}. The focus of much current research in cosmology is to understand the properties and origins of the dark sector, in particular dark energy, for which the cosmological constant is the simplest explanation \cite{Einstein:1917ce}; this standard model is currently in very good agreement with current cosmological observations. Theoretically, however, the coincidence and fine--tuning problems challenge our understanding of gravity and quantum field theory \cite{Weinberg:1988cp,Zlatev:1998tr}. A plethora of alternative dynamical dark energy models have been proposed, such as quintessence \cite{Wetterich:1987fm,Peccei:1987mm,Peebles:1987ek}, k-essence \cite{Chiba:1999ka,ArmendarizPicon:2000dh}, phantom \cite{Caldwell:1999ew}, Chaplygin gas \cite{Bento:2002ps}, Ricci dark energy \cite{Gao:2007ep}, and holographic dark energy and related ideas \cite{Li:2004rb,Cohen:1998zx}. Furthermore, coupled dark energy models have also been extensively studied since, from the field theoretic point of view, dark energy is not prohibited from interacting with cold dark matter \cite{Wetterich:1994bg,Amendola:1999er,Wintergerst:2010ui,Pettorino:2008ez,Mangano:2002gg,Amendola:2003wa,Koivisto:2005nr,Koivisto1,Guo:2007zk,Quercellini:2008vh,Quartin:2008px,Valiviita:2009nu,Jack} or, for example, massive neutrinos \cite{Gu:2003er,Fardon:2003eh,Brookfield:2005bz,Ichiki:2008rh,Wetterich:2007kr}. 

In this paper we consider the case of a (non--universally) coupled dark energy model in which dark matter particles feel an additional fifth force mediated by the dark energy scalar field. This coupling between the dark sector elements modifies the background evolution of the Universe, as well as the growth of perturbations: in this paper we concentrate on constraints coming from the background only, deferring the perturbed case for future work. As conformally coupled dark matter models have been well studied \cite{Amendola:2003eq,Xia:2009zzb,Pettorino:2012ts,Amendola:2000ub,Bean:2008ac,Pettorino:2013oxa,Xia:2013nua,Ade:2015rim,Amendola:2011ie}, and tight constraints on the model parameters have been established \cite{Pettorino:2013oxa,Xia:2013nua,Ade:2015rim}, the main aim of this paper will be to augment the models of these studies with a disformal coupling and discern its influence in light of the conformal-only constraints. Models that utilise such disformal interactions within the dark sector have been attracting much attention recently \cite{Zuma5,Zumalacarregui:2012us,Jack,Koivisto,Koivisto1,Sakstein:2014isa,Carsten,Sakstein:2014aca,vandeBruck:2015rma,vandeBruck:2016jgg}, so it has become an imperative that they be compared with state-of-the-art cosmological data sets. 

This paper is structured as follows. In Section \ref{sec:Model} we introduce our coupled dark energy model and present the background evolution equations in a flat, homogeneous, and isotropic Universe. We list in Section \ref{sec:data} the observational data sets we will use here to derive constraints on our model parameters, while in Section \ref{sec:results} we present the obtained constraints for each coupled dark matter model. Finally Section \ref{sec:conclusions} contains our conclusions, and outlines future work.

\section{Theoretical model: action \& equations of motion}
\label{sec:Model}
We consider the scalar--tensor theory described by the following action, expressed in the Einstein frame:
\begin{equation}\label{action}
\mathcal{S} = \int d^4 x \sqrt{-g} \left[ \frac{M_{\rm Pl}^2}{2} R - \frac{1}{2} g^{\mu\nu}\partial_\mu \phi\, \partial_\nu \phi - V(\phi) + \mathcal{L}_{SM}\right] + \int d^4 x \sqrt{-\tilde{g}} \mathcal{\tilde{L}}_{DM}\left(\tilde g_{\mu\nu}, \psi\right),
\end{equation}
where $\kappa^2\equiv M_\text{Pl}^{-2}\equiv 8\pi G$ such that $M_\text{Pl}=2.4\times 10^{18}$ GeV is the reduced Planck mass, dark energy is described by a quintessence scalar field, $\phi$, with a potential, $V(\phi)$, and the uncoupled standard model (SM) particles are described by the Lagrangian, $\mathcal{L}_{SM}$, which includes a relativistic component, $r$, and a baryon component, $b$. Particle quanta of the dark matter fields, $\psi$, propagate on geodesics defined by the metric
\begin{equation}\label{disformal_relation}
\tilde g_{\mu\nu} = C(\phi) g_{\mu\nu} + D(\phi)\, \partial_\mu\phi\, \partial_\nu \phi\;, 
\end{equation}
with $C(\phi),\;D(\phi)$ being the conformal and disformal coupling functions respectively. In the general case, the free functions $C$ and $D$ can depend on the kinetic term $X = -\frac{1}{2} g^{\mu\nu}\partial_\mu \phi \partial_\nu \phi$ as well, but throughout this paper we will not consider such a scenario. By definition, in the Einstein frame the gravitational sector has the Einstein--Hilbert form, and SM particles are not coupled to the scalar field directly. 

The action above defines an interaction between dark matter and dark energy, resulting from the modification of the gravitational field experienced by the dark matter particles, $\tilde{g}_{\mu\nu}$, by the dark energy scalar field.

Variation of the action (\ref{action}) with respect to the metric $g_{\mu\nu}$ leads to the field equations
\begin{equation}\label{EFE}
R_{\mu\nu}\,-\,\frac{1}{2}g_{\mu\nu}R = \kappa^2\left(T^\phi_{\mu\nu} + T^{SM}_{\mu\nu} + T^{DM}_{\mu\nu}\right)\;,
\end{equation}
where the energy--momentum tensors of the scalar field, SM particles, and dark matter particles are defined by 
\begin{eqnarray}
T^\phi_{\mu\nu}&=&\partial_\mu \phi \partial_\nu \phi\,-\,g_{\mu\nu}\left(\frac{1}{2}g^{\rho\sigma}\partial_\rho\phi\partial_\sigma\phi\;\;+\;\;V(\phi)\right)~,\nonumber\\
T^{SM}_{\mu\nu}&=&-\frac{2}{\sqrt{-g}} \frac{\delta\bigl(\sqrt{-g}\mathcal{L}_{SM}\bigr)}{\delta g^{\mu\nu}}\;,\;
T^{DM}_{\mu\nu}=-\frac{2}{\sqrt{-g}} \frac{\delta\bigl(\sqrt{-\tilde{g}}\tilde{\mathcal{L}}_{DM}\bigr)}{\delta g^{\mu\nu}}\;,  \nonumber
\end{eqnarray}
respectively. Non-conservation of $T^{\phi}_{\mu\nu}$ implies the following relation
\begin{equation}\label{eq:modKG}
\Box\phi=V_{,\phi} - Q\;,
\end{equation}
where
\begin{equation}\label{coupling}
Q=\frac{C_{,\phi}}{2C}T_{DM}+\frac{D_{,\phi}}{2C}T_{DM}^{\mu\nu}\nabla_\mu\phi\nabla_\nu\phi-\nabla_\mu\left[\frac{D}{C}T^{\mu\nu}_{DM}\nabla_\nu\phi\right]\;,
\end{equation}
and $T_{DM}$ is the trace of $T_{DM}^{\mu\nu}$, which satisfies a modified conservation equation  
\begin{equation}\label{cons}
\nabla^\mu T^{DM}_{\mu\nu}=Q\nabla_\nu\phi\;.
\end{equation}
Since SM particles are uncoupled from the scalar field, their energy--momentum tensor obeys the standard conservation equation
\begin{equation}
\nabla^\mu T^{SM}_{\mu\nu}=0\;.
\end{equation}
We assume all species to be perfect fluids:
\begin{equation}
T^{\mu\nu}_{i}=(\rho_{i}+p_{i})u^\mu u^\nu+p_{i} g^{\mu\nu}\;,
\end{equation}
where the index $i$ stands for dark matter and SM. The Einstein frame SM and DM fluid's energy density and pressure are denoted by $\rho_{i}$ and $p_{i}$ respectively. 

As we state in the introduction, only the background dynamics of the theory are considered in this work---a study of the perturbations will appear in a future publication and so from now on we will consider the standard flat Friedmann-Robertson-Walker (FRW) metric, given by
\begin{equation}
ds^2 = g_{\mu\nu}dx^{\mu} dx^{\nu} = a^2(\tau)\left[-d\tau^2 + \delta_{ij} dx^i dx^j\right]\;,
\end{equation} 
with conformal time $\tau$, we will denote a conformal time derivative by a prime, and scale factor $a(\tau)$. Spatial gradients in the scalar field, $\phi$, and matter fluid variables, $\rho_i$, $p_i$, are hence also neglected for this first paper.

Given the above simplifications, the modified Klein-Gordon equation, \eqref{eq:modKG}, becomes
\begin{equation}
\phi^{\prime\prime} + 2 \mathcal{H} \phi^{\prime} + a^2 V_{,\phi} = a^2 Q\;,
\end{equation}
the fluid conservation equations simplify to
\begin{eqnarray}
\rho_r^\prime + 4\mathcal{H}\rho_r &=& 0\;, \\
\rho_b^\prime + 3\mathcal{H}\rho_b &=& 0\;, \\
\rho_c^\prime + 3\mathcal{H}\rho_c &=& -Q\phi^{\prime}\;,
\end{eqnarray}
and the Friedmann equations to
\begin{eqnarray}
\mathcal{H}^2 &=& \frac{\kappa^2}{3}a^2\left(\rho_\phi + \rho_b + \rho_r + \rho_c \right), \\
\mathcal{H}^\prime &=& -\frac{\kappa^2}{6} a^2 \left(\rho_\phi + 3p_{\phi} + \rho_b + 2\rho_r + \rho_c\right),
\end{eqnarray}
where we now denote coupled DM by a subscript, $c$. The scalar field's energy density and pressure respectively have the usual forms: $\rho_\phi={\phi^\prime}^2/\left(2a^2\right)+V(\phi)$, $p_\phi=\rho_\phi-2V(\phi)$, and the conformal Hubble parameter we denote $\mathcal{H}=a^\prime/a$.
The coupling as defined by equation (\ref{coupling}) simplifies to \cite{Carsten}
\begin{equation}\label{Q}
Q=-\frac{a^2C_{,\phi} + D_{,\phi}{\phi^\prime}^2 - 2D\left(\frac{C_{,\phi}}{C}{\phi^\prime}^2 + a^2V_{,\phi} + 3\mathcal{H}\phi^\prime\right)}{2\left[a^2C + D\left(a^2\rho_c - {\phi^\prime}^2\right)\right]} \rho_c \;.
\end{equation}
%
Throughout this paper we choose an exponential scalar field potential
\begin{equation}
V(\phi)=V_0^4 e^{-\lambda\kappa\phi}\;,
\end{equation}
where $V_0$ and $\lambda$ are constants. When we consider a conformal coupling, we make use of an exponential function
\begin{equation}
C(\phi) = e^{2\alpha\kappa\phi}\;,
\end{equation}
where $\alpha$ is a constant. As this is a simple first study, we only take into account a constant disformal coupling
\begin{equation}\label{disformal_function}
D(\phi)=D_M^4\;,
\end{equation}
where $D_M$ is a constant inverse mass scale, expressed in $\text{meV}^{-1}$. 

%
%

\begin{figure}[t]
\centering
  \includegraphics[width=0.49\textwidth]{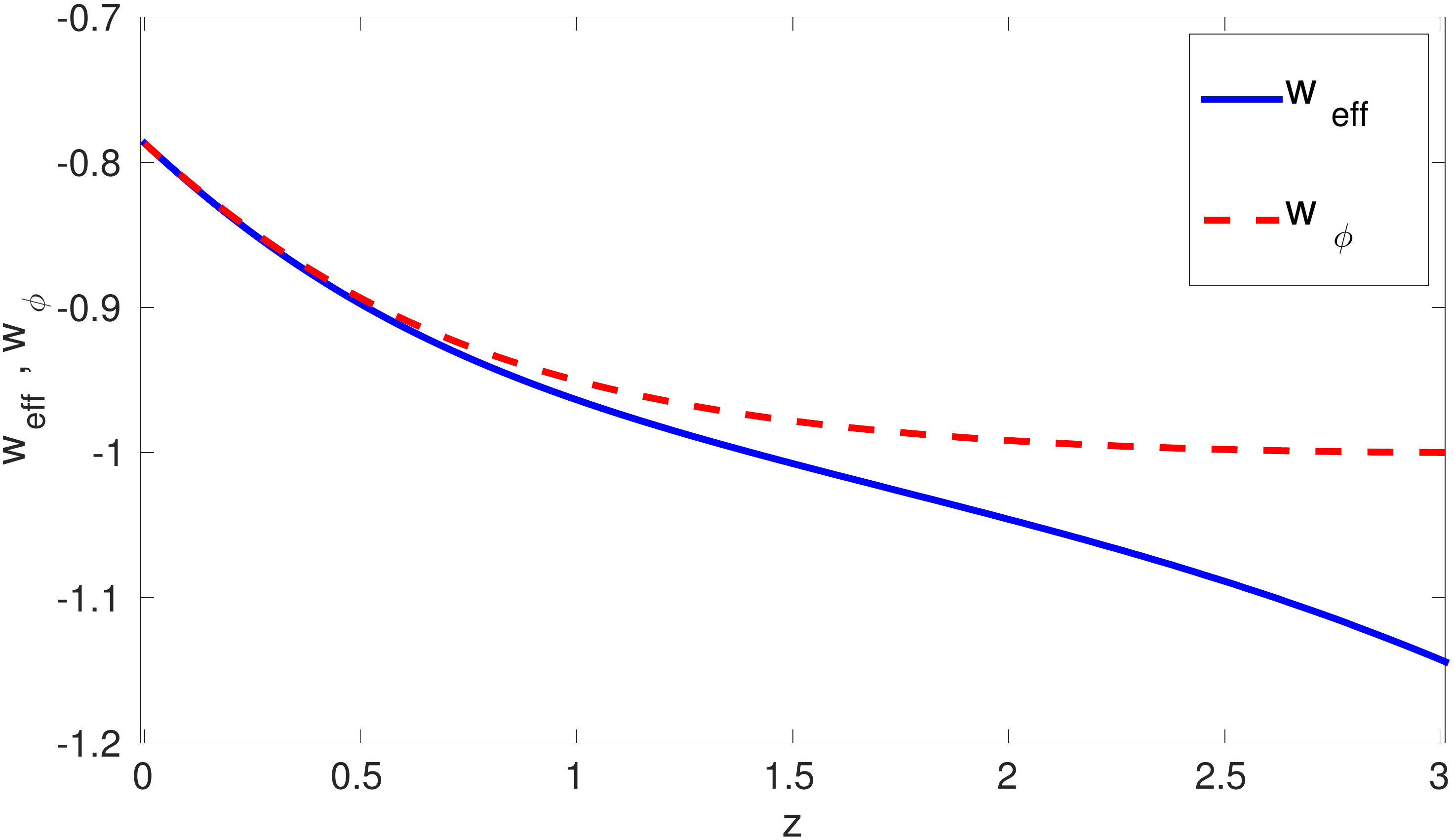}
  \includegraphics[width=0.49\textwidth]{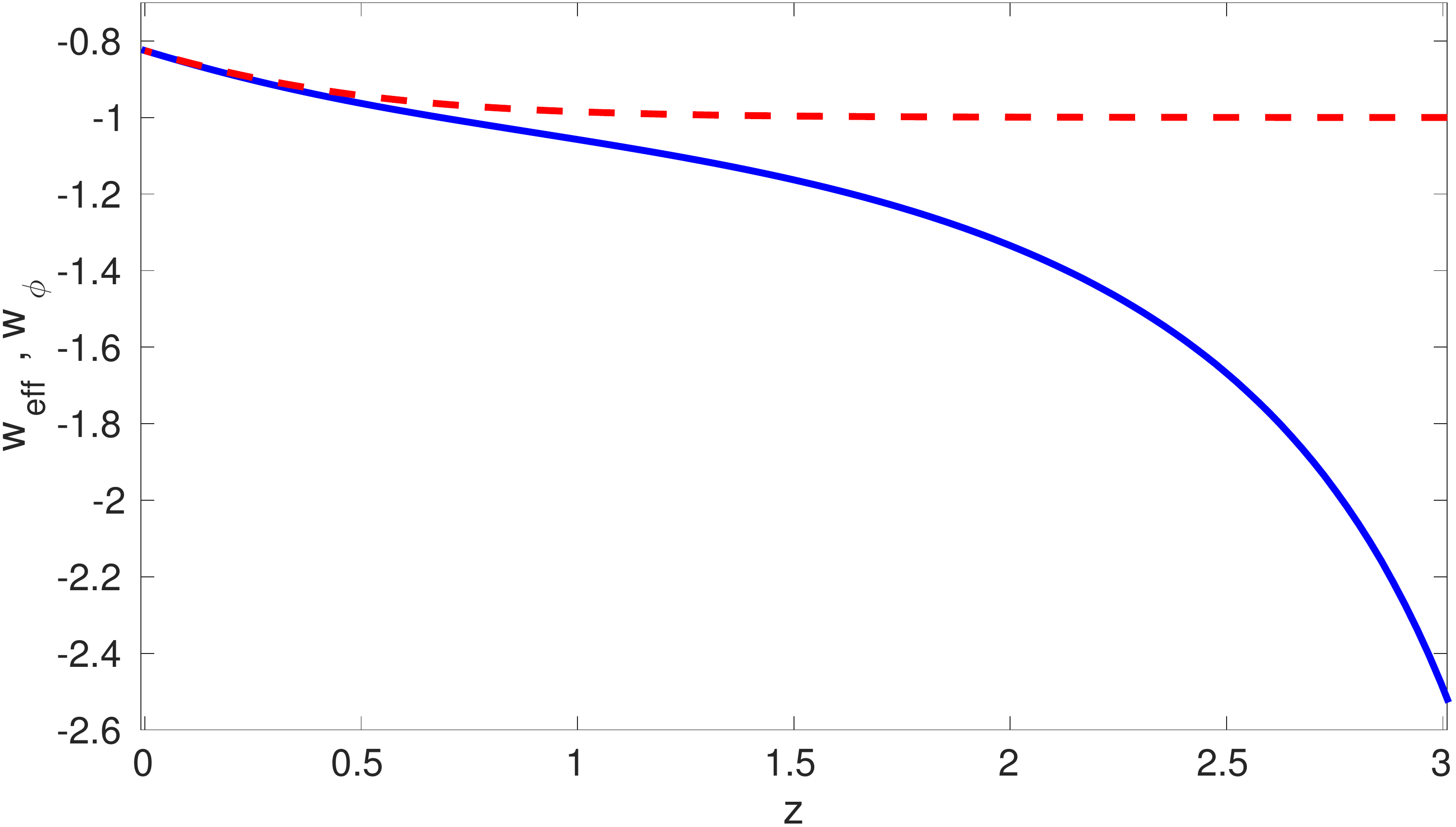}
  \includegraphics[width=0.49\textwidth]{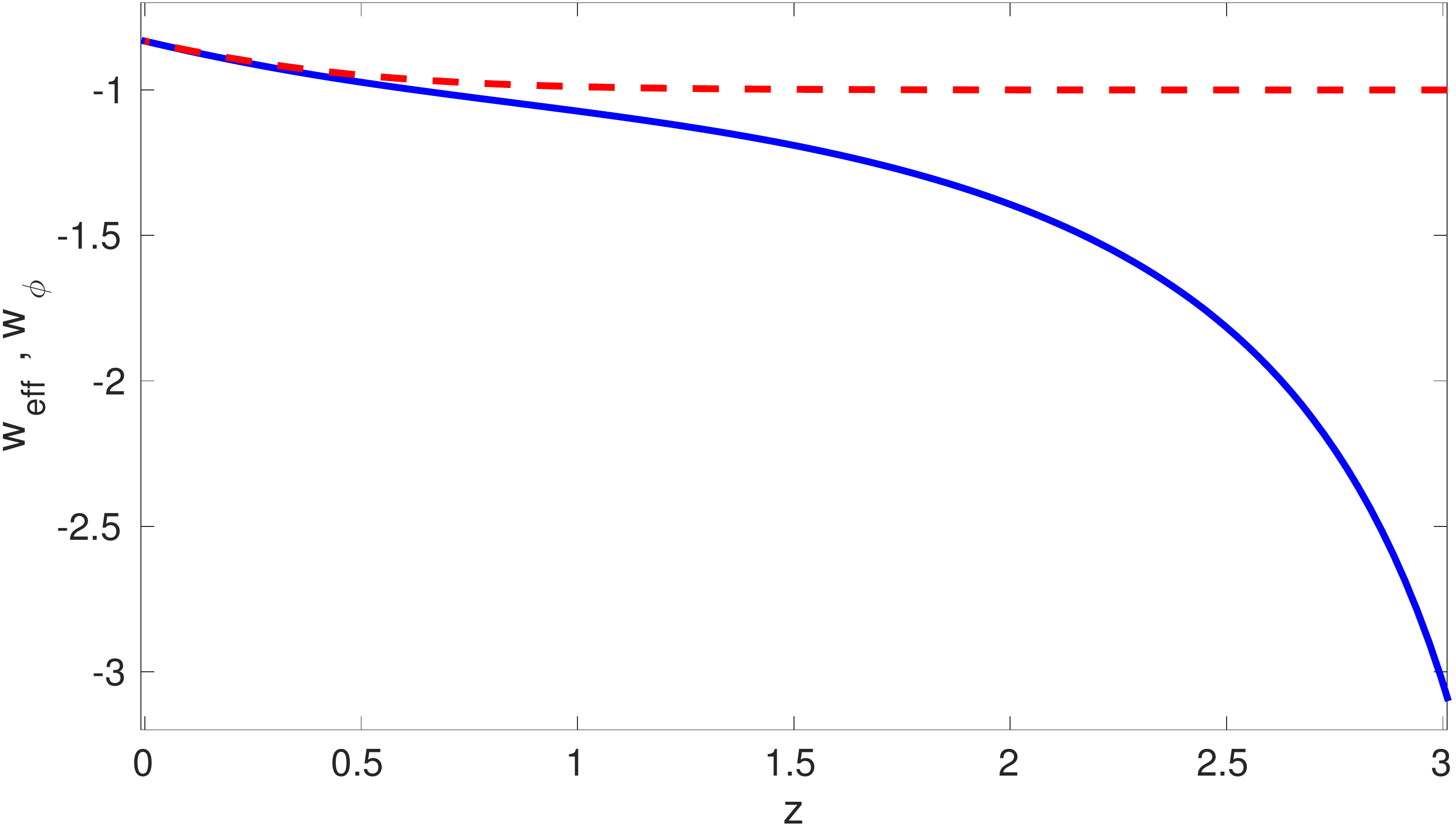}
 \caption{These figures show the evolution of the effective equation of state (solid) and the corresponding evolution of the scalar field equation of state parameter (dashed). We show a conformal case with $\alpha=0.02$ (left), a disformal case with $D_M=0.34\,\text{meV}^{-1}$ (right), and a conformal disformal case with $\alpha=0.02$ and $D_M=0.34\, \text{meV}^{-1}$ (bottom). In all cases we set $\lambda=1.2$.  } 
\label{fig:couplings}
\end{figure}

Let us now consider a phenomenological re-parameterisation of the system made concrete above. We will find interpretation of our parameter constraints in the following sections is made much more clear if we re-parameterise the system described above in the following way, and we will return to comment on these definitions with regards to our results in later sections. Following Ref. \cite{Jack,Khoury,vandeBruck:2015rma,Basilakos:2013vya}, we repackage the dark sector of our model by now defining an effective dark energy fluid, $\rho_{\text{DE,eff}}$, with effective equation of state, $w_{\text{eff}}(z)$, such that:
\begin{equation}
\rho_{\text{DE,eff}}^\prime + 3\mathcal{H}\left(1+w_{\text{eff}}\right)\rho_{\text{DE,eff}}=0\;,
\end{equation}
and
\begin{equation}
\rho_{\text{c,eff}}^\prime + 3\mathcal{H}\rho_{\text{c,eff}}=0\;,
\end{equation}
hence
\begin{equation}
\mathcal{H}^2=\frac{\kappa^2}{3} a^2\left(\rho_{\text{DE,eff}}+\rho_b+\rho_r+\rho_{c,0}a^{-3}\right)\;.
\end{equation}
In this re-parameterised system there are by definition no dark sector interactions, and the dark matter energy density dilutes with the expansion as $a^{-3}$.
By comparing these non--interacting dark sector definitions with our coupled dark energy model equations, we get that 
\begin{equation}\label{eff}
w_{\text{eff}}=\frac{p_\phi}{\rho_{\text{DE,eff}}}=\frac{p_\phi}{\rho_\phi + \rho_c - \rho_{c,0}a^{-3}}\;.
\end{equation}
Since the coupled DM energy density does not redshift as $a^{-3}$, it follows that although $w_\phi\in[-1,1]$, $w_{\text{eff}}$ can take values less than -1. We have defined $w_{\text{eff}}$ in equation \eqref{eff} above such that, evaluated today, the effective equation of state coincides with the scalar field equation of state parameter. We illustrate the evolution of the effective equation of state and the scalar field equation of state parameter in Fig. \ref{fig:couplings} for three different coupling scenarios.   
%

\section{Observational data sets}
\label{sec:data}
For our main analysis presented in Section \ref{sec:results} we shall be considering constraints on the cosmological parameters derived from the late-time Universe expansion history. We shall be considering Hubble parameter measurements \cite{Farooq:2013hq}, BAO data \cite{Cuesta:2015mqa,Beutler:2011hx,Ross:2014qpa}, together with SNIa data from the Union2.1 catalogue \cite{Suzuki:2011hu}. Moreover, we shall be considering a standard big bang nucleosynthesis (BBN) prior corresponding to a baryon density $100\Omega_{b}h^2=2.202\pm0.046$ \cite{Cooke:2013cba}.

\subsection{$H(z)$ data set and the Hubble constant}
\label{sec:Hubble}
We use $H(z)$ data inferred from the differential age technique \cite{Jimenez:2001gg}, a technique based on measurements of the age difference between two passively--evolving galaxies that formed at the same time but are separated by a small redshift interval, i.e. a measurement of the derivative $dz/dt$, where $t$ is the cosmic time and $H=a^{-1}\mathcal{H}$. 
In Section \ref{sec:results}, we use 28 independent $H(z)$ measurements \cite{Farooq:2013hq}, between redshifts $0.07\leq z\leq 2.3$ to place constraints on our model parameters. We also consider a Gaussian prior on the Hubble constant\footnote{We are aware of a more recent measurement of the Hubble constant as reported in Ref. \cite{Riess:2016jrr}, although we decided to use a more conservative constraint in our analysis.}, given by the Hubble Space Telescope (HST) measurement of $H_0 =73.8\pm 2.4 \,\text{km}\,\text{s}^{-1}\,\text{Mpc}^{-1}$ \cite{Riess:2011yx}.

\subsection{Baryon Acoustic Oscillations}
\label{sec:BAO}
BAO features in the clustering of galaxies are being used by large scale surveys as a standard ruler to measure the distance--redshift relation. The acoustic oscillations in the photon--baryon plasma arise from the tight coupling of baryons and photons in the radiation era. BAO data is usually reported in terms of the angle--averaged distance
\begin{equation}
D_V(z)=\left[z(1+z)^2D_A^2(z)H^{-1}(z)\right]^{1/3}\;,
\end{equation}
consisting of the angular diameter distance, $D_A(z)$, and the Hubble parameter. 
In the main analysis of Section \ref{sec:results} we use the CMASS and LOWZ samples from Data Release 12 of the Baryon Oscillation Spectroscopic Survey (BOSS) at $z_{\text{eff}} = 0.57$ and $z_{\text{eff}} = 0.32$ respectively \cite{Cuesta:2015mqa}, the 6dF Galaxy Survey (6dFGS) measurement at $z_{\text{eff}} = 0.106$ \cite{Beutler:2011hx}, and the Main Galaxy Sample of Data Release 7 of Sloan Digital Sky Survey (SDSS-MGS) at $z_{\text{eff}} = 0.15$ \cite{Ross:2014qpa}.

\subsection{Type Ia Supernovae}
\label{sec:SNIa}
\begin{figure}
\centering
  \includegraphics[width=1\textwidth]{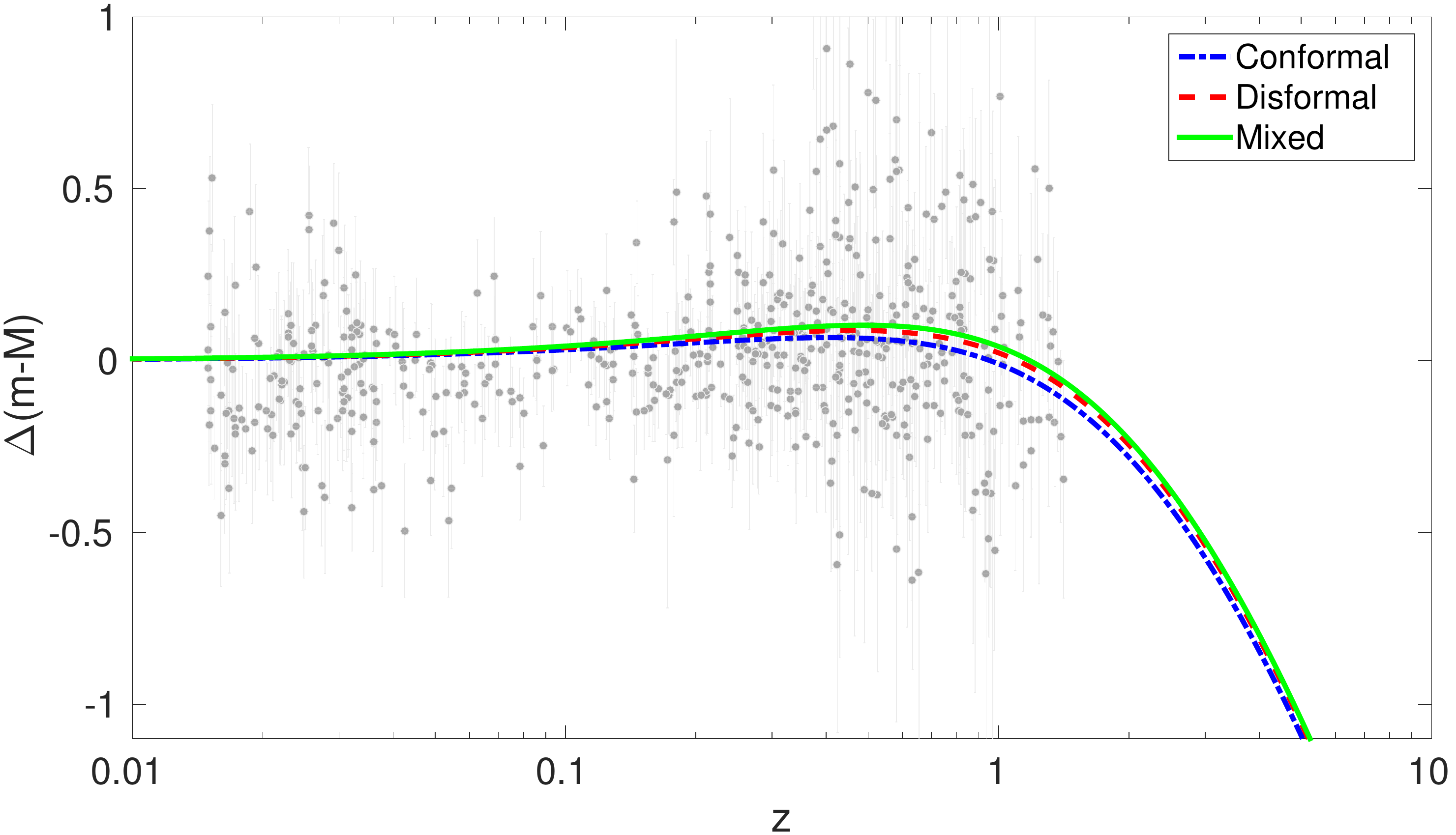}
  \caption{In this figure we show the distance modulus for three different models together with the supernova Union2.1 data set \cite{Suzuki:2011hu}. We illustrate a conformal case with $\alpha=0.02$, a disformal case with $D_M=0.4\,\text{meV}^{-1}$, and a mixed conformal disformal case with $\alpha=0.18$ and $D_M=0.4\, \text{meV}^{-1}$. In all cases we set $\lambda=1.1$.  } 
\label{fig:SN}
\end{figure}

Apart from providing observational evidence for the accelerating expansion of the Universe \cite{Perlmutter:1998np,Riess:1998cb,Garnavich:1998th,Schmidt:1998ys,Perlmutter:1997zf}, SNIa observations have also been widely used for cosmological model parameter--fitting. In our analysis we use the supernova Union2.1 compilation of 580 data points \cite{Suzuki:2011hu}. In Fig. \ref{fig:SN} we show the residual Hubble diagram from an empty Universe, for three classes of models compared to the data set of Ref. \cite{Suzuki:2011hu}. The distance modulus is defined as \cite{Amendola:2004ew}
\begin{equation}
\Delta(m-M)=(m-M)_{\text{model}}-(m-M)_{\text{Milne}}\;,\;\;m-M=5\log_{10}\frac{D_L(z)}{10\,\text{pc}}\;,
\end{equation} 
where $m$ is the apparent magnitude, $M$ is the absolute magnitude of the object, and $D_L(z)$ is the luminosity distance. 

\section{Parameter constraints \& best fit values}
\label{sec:results}
For the global fitting of the cosmological parameters, we use a modified version of the CLASS code \cite{Blas:2011rf} to evolve the coupled dark energy--dark matter background equations, and interface with the public (Metropolis--Hastings) Markov chain Monte Carlo (MCMC) code Monte Python \cite{Audren:2012wb} to constrain the model parameter space with cosmological data. The amplitude of the scalar field exponential potential function, $V_0$, is determined by using an iterative routine in the modified CLASS code. We assume top--hat priors for our parameters: the baryon energy density parameter $\Omega_b h^2\in[0.005,0.1]$, the coupled cold dark matter energy density parameter $\Omega_c h^2\in[0.01,0.99]$, the Hubble parameter $H_0\in[45,90]\,\text{km}\,\text{s}^{-1}\,\text{Mpc}^{-1}$, the conformal coupling parameter $\alpha\in[0,0.48]$, the disformal coupling parameter $D_M\in[0,1.1]\,\text{meV}^{-1}$, and the scalar field potential exponent parameter $\lambda\in[0,1.7]$. On top of these, we also include Gaussian priors on $\Omega_b h^2$ and $H_0$, as mentioned in Sections \ref{sec:data} and \ref{sec:Hubble}. Following the dynamical systems analysis in Ref. \cite{vandeBruck:2016jgg}, we have chosen the range for our model parameters $\alpha,\,D_M,$ and $\lambda$ to accommodate all the values for which there is acceleration at the present. Although in this paper we shall only consider positive values for our parameters, we have repeated the analysis presented below for a negative range of priors and the obtained results were consistent with those presented here. Changing the scalar field's initial value, $\phi_{\mathrm{ini}}$, is equivalent to changing the field potential height parameter $V_0$, so we have held $\phi_{\mathrm {ini}}$ fixed for the entire study. 
\begin{table}
\begin{center}
\begin{tabular}{ c  c  c  c } 
 \hline
\hline
 Parameter~  &  ~$H$+BAO+SNIa~ & ~$H$+BAO+SNIa+BBN~ & ~$H$+BAO+SNIa+BBN+HST~  \\ 
\hline
$\Omega_b h^2$ & $0.021^{+0.0072}_{-0.0069}$ & $0.022^{+0.0005}_{-0.0005}$ & $0.022^{+0.0005}_{-0.0005}$  \\
$\Omega_c h^2$ & $0.11^{+0.013}_{-0.011}$ & $0.11^{+0.007}_{-0.007}$ & $0.11^{+0.008}_{-0.008}$ \\
$H_0$ & $67.49^{+2.14}_{-2.18}$ & $67.92^{+1.47}_{-1.57}$ & $70.14^{+1.35}_{-1.63}$ \\
$\lambda$ & $<1.27$ & $<1.21$ & $<1.05$ \\
$\alpha$ & $<0.193$ &  $<0.143$ & $0.097^{+0.056}_{-0.039}\;(<0.168)$ \\  
\hline
\hline
\end{tabular}
\end{center}
\caption{\label{table:conformal} For each model parameter we report the best fit values and $1\sigma$ errors in the conformally coupled DM scenario. For $\lambda$ and $\alpha$ we quote the $95.4\%$ upper limits instead. See the top of Section \ref{sec:results} for our chosen parameter priors. In the HST run we further include the best fit value and $1\sigma$ errors for the conformal coupling strength parameter. }
\end{table}
\begin{figure}
\centering
  \includegraphics[width=1\textwidth]{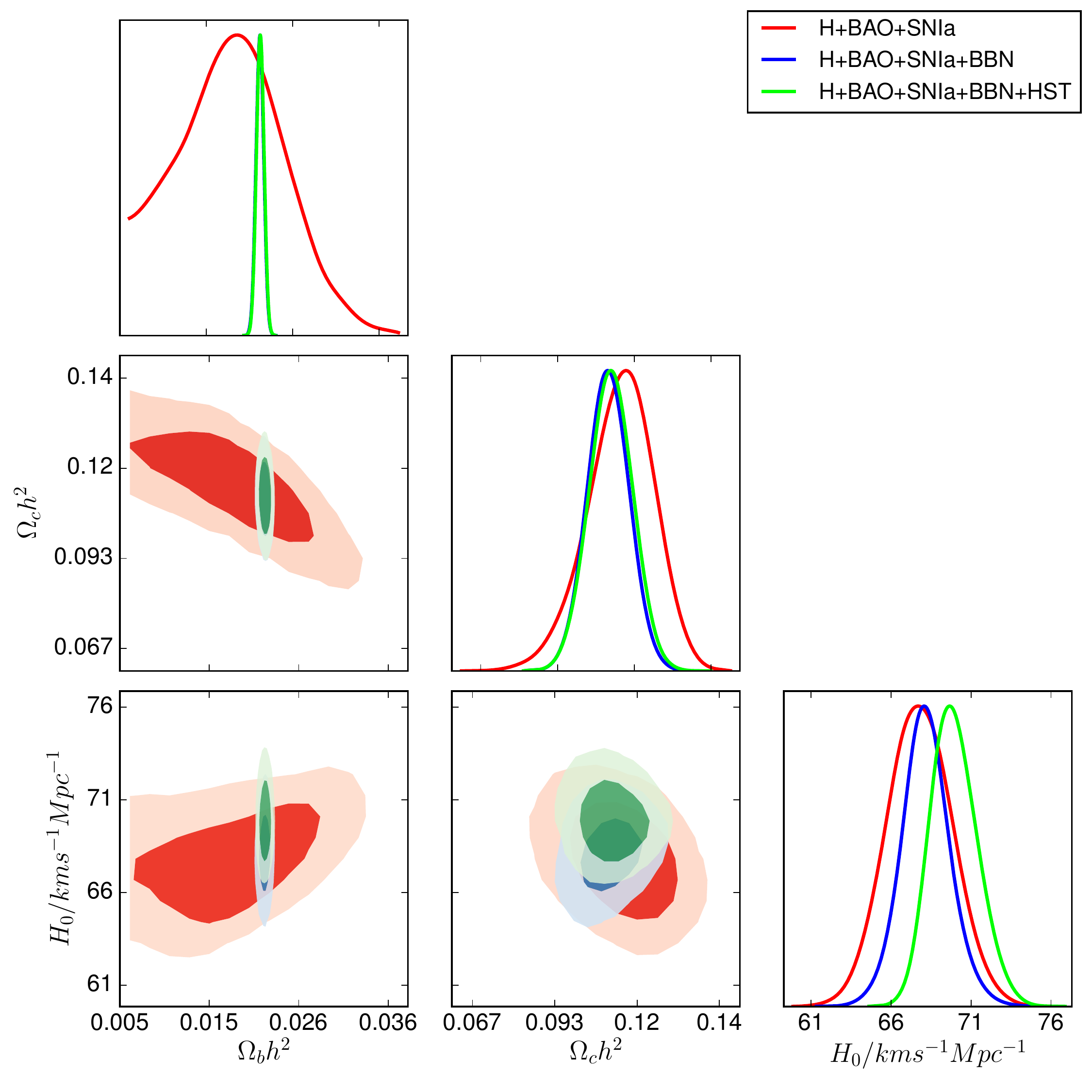}
\caption{Confidence--level contours of the cosmological parameters for the conformally coupled DM case. We compare the $68.3\%$ (dark shaded) and $95.4\%$ (light shaded) constraints arising from $H+$SNIa+BAO observations with $H+$SNIa+BAO+BBN and $H+$SNIa+BAO+BBN+HST observations. The marginalized one--dimensional posterior distributions are also shown for comparison. } 
\label{fig:conformal_cosmo}
\end{figure}
\subsection{Conformal case}
\label{sec:conformal}

\begin{figure}
\centering
  \includegraphics[width=0.7\textwidth]{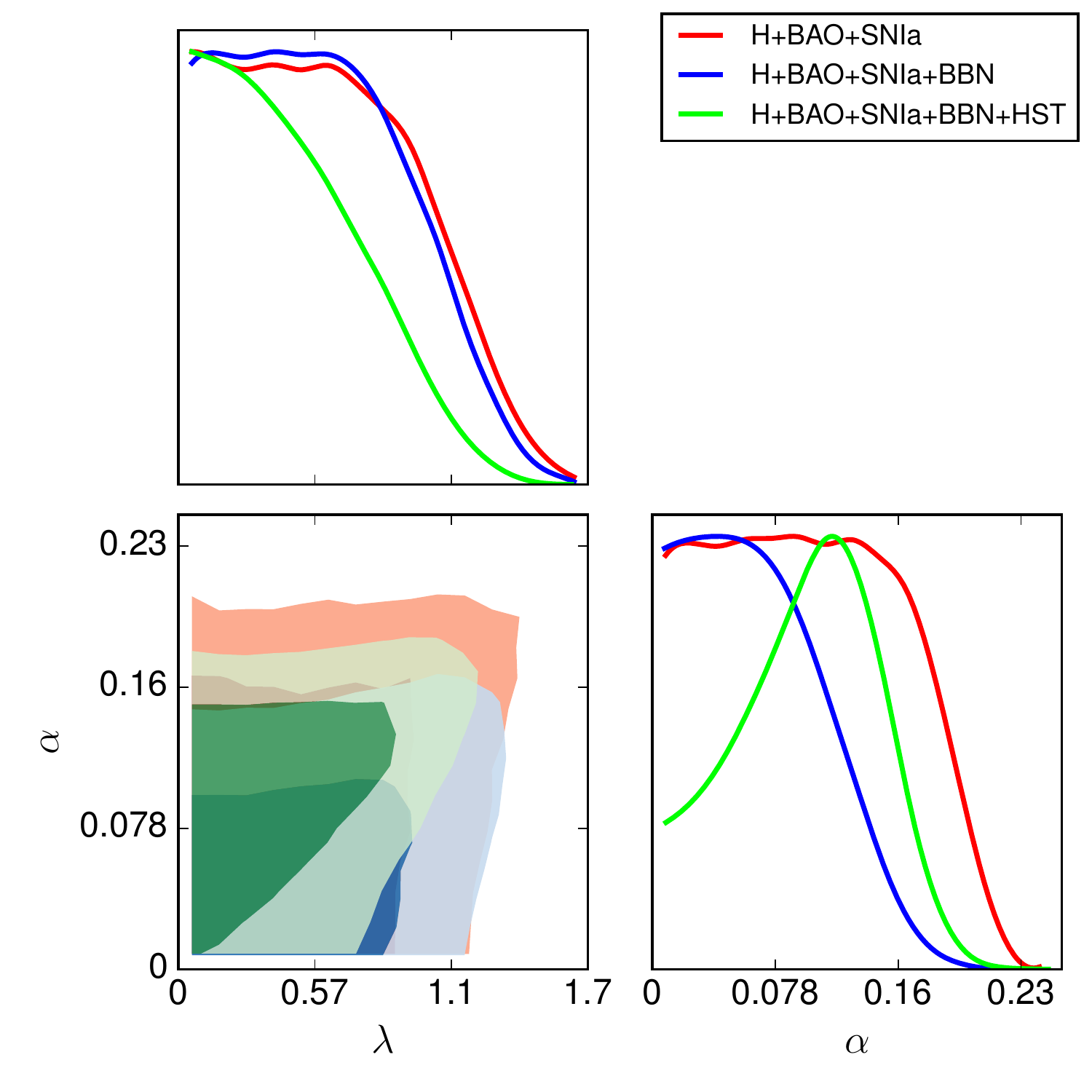}
\caption{Confidence--level contours of the model parameters for the conformally coupled DM case. We compare the $68.3\%$ (dark shaded) and $95.4\%$ (light shaded) constraints arising from $H+$SNIa+BAO observations with $H+$SNIa+BAO+BBN and $H+$SNIa+BAO+BBN+HST observations. The marginalized one--dimensional posterior distributions are also shown for comparison. } 
\label{fig:conformal_alpha_lambda}
\end{figure}

We first discuss the well known case in which dark matter is only conformally coupled \cite{Amendola:2003eq,Xia:2009zzb,Pettorino:2012ts,Amendola:2000ub,Bean:2008ac,Pettorino:2013oxa,Xia:2013nua,Ade:2015rim,Amendola:2011ie}. Although already well documented, this case is presented here both as a consistency check and to provide the means to cleanly compare parameter constraints derived from the purely conformal case with the mixed case discussed in Section \ref{sec:conformal_disformal}. Our results from different runs of Monte Python are illustrated in Table \ref{table:conformal}. The confidence--level contours and the corresponding one--dimensional posterior distributions for the $H$+BAO+SNIa (red contours) run, the $H$+BAO+SNIa+BBN (blue contours) run and the $H$+BAO+SNIa+BBN+HST (green contours) run are shown in Fig. \ref{fig:conformal_cosmo} and Fig. \ref{fig:conformal_alpha_lambda}. Using the $H$+BAO+SNIa observations, we obtain an upper limit on the interaction coupling strength $\alpha<0.193$ at the $95.4\%$ confidence level (c.l.).

When we include the BBN prior on the baryon energy density parameter, the upper limit on the conformal coupling parameter improves slightly to $\alpha<0.143$ at the $95.4\%$ c.l., which is mainly due to better constraints on the cosmological parameters. The obtained upper limit is consistent with other results in the literature \cite{Amendola:2003eq,Xia:2009zzb,Pettorino:2012ts,Amendola:2000ub,Bean:2008ac,Pettorino:2013oxa,Xia:2013nua,Ade:2015rim}. When using the HST prior in combination with the other data sets, the conformal coupling strength parameter upper limit increases, as expected \cite{Pettorino:2012ts,Pettorino:2013oxa,Xia:2013nua}, to $\alpha<0.168$ $(95.4\%\;\text{c.l.})$. Indeed, we find that the best fit value for the conformal coupling strength is away from zero at $1\sigma$, $\alpha=0.097^{+0.056}_{-0.039}$, but is consistent with zero at $2\sigma$. This occurs mainly due to a slight tension between different values of $H_0$ deduced from the data sets. In this model, the potential slope $\lambda$ is constrained to be $\lambda<1.21$ $(95.4\%\;\text{c.l.})$ without the HST data, and $\lambda<1.05$ $(95.4\%\;\text{c.l.})$ when including the HST measurement; both are consistent with results in the literature \cite{Xia:2009zzb,Bean:2008ac}. The data we use in our analysis is not able to tightly constrain the conformal coupling interaction parameter very well; tighter constraints have been obtained when using recent CMB data \cite{Pettorino:2013oxa,Xia:2013nua,Ade:2015rim}. 

%
%
%

%
\begin{table}[t]
\begin{center}
\begin{tabular}{ c  c  c  c } 
 \hline
\hline
 Parameter~   &  ~$H$+BAO+SNIa~ & ~$H$+BAO+SNIa+BBN~ & ~$H$+BAO+SNIa+BBN+HST~  \\ 
\hline
$\Omega_b h^2$ & $0.021^{+0.0046}_{-0.0053}$ & $0.022^{+0.0005}_{-0.0005}$ & $0.022^{+0.0005}_{-0.0005}$ \\
$\Omega_c h^2$ & $0.11^{+0.013}_{-0.011}$ & $0.11^{+0.008}_{-0.008}$ & $0.11^{+0.007}_{-0.008}$ \\
$H_0$ & $67.57^{+2.19}_{-2.24}$ & $67.79^{+1.22}_{-1.11}$ & $68.53^{+0.95}_{-0.92}$ \\
$\lambda$ & $<1.56$ &  $<1.56$ & $<1.53$ \\
$D_M$ & $>0.070$ &  $>0.074$ & $>0.094$ \\  
\hline
\hline
\end{tabular}
\end{center}
\caption{\label{table:disformal} For each cosmological parameter we report the best fit values and $1\sigma$ errors in the disformally coupled DM scenario. For $\lambda$ and $D_M\;\left(\text{meV}^{-1}\right)$, we quote the $95.4\%$ limits instead. See the top of Section \ref{sec:results} for the parameter priors.}
\end{table}
%
\subsection{Disformal case}
\label{sec:disformal}
We now discuss the constraints on the purely disformal coupled case, in which the dark matter and dark energy are interacting via a constant disformal coupling as defined in (\ref{disformal_function}) with $C(\phi)=1$. From our choice of data sets we deduce that a non--zero constant disformal coupling is preferred above a $2\sigma$ confidence level. When using the $H$+BAO+SNIa data we observe that $D_M>0.070\;\text{meV}^{-1}$ $(95.4\%\;\text{c.l.})$, and when combining this data with the BBN prior we get that $D_M>0.074\;\text{meV}^{-1}$ $(95.4\%\;\text{c.l.})$. The obtained limits are given in Table \ref{table:disformal}. This non--zero coupling preference distinguishes the purely disformal coupling from the purely conformal coupling, although we should remark that a non--zero conformal coupling was also found to be slightly favoured particularly when combining astrophysical data sets \cite{Xia:2013nua,Pettorino:2013oxa,Ade:2015rim}. In the purely conformal case the peak away from zero, which we discussed in Section \ref{sec:conformal}, and was also reported in Ref. \cite{Xia:2013nua,Pettorino:2013oxa,Ade:2015rim}, is still not pronounced enough to claim evidence for a deviation away from the concordance model has been found. This is due to a number of possible systematics. 

\begin{figure}
\centering
  \includegraphics[width=1\textwidth]{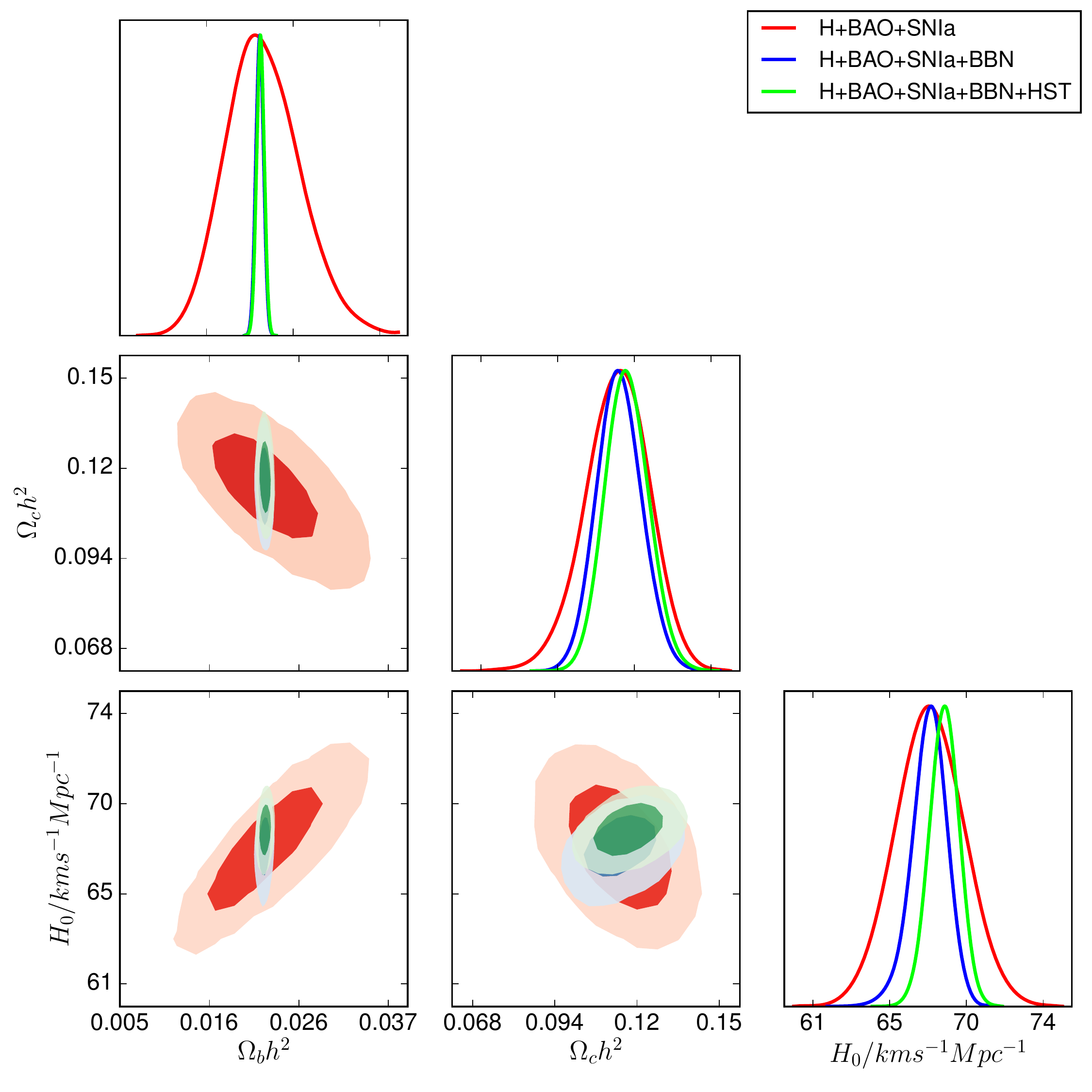}
  \caption{Confidence--level contours of the cosmological parameters for the disformally coupled DM case. We compare the $68.3\%$ (dark shaded) and $95.4\%$ (light shaded) constraints arising from $H+$SNIa+BAO observations with $H+$SNIa+BAO+BBN and $H+$SNIa+BAO+BBN+HST observations. The marginalized one--dimensional posterior distributions are also shown for comparison. } 
\label{fig:disformal_cosmo}
\end{figure}

\begin{figure}
\centering
  \includegraphics[width=0.7\textwidth]{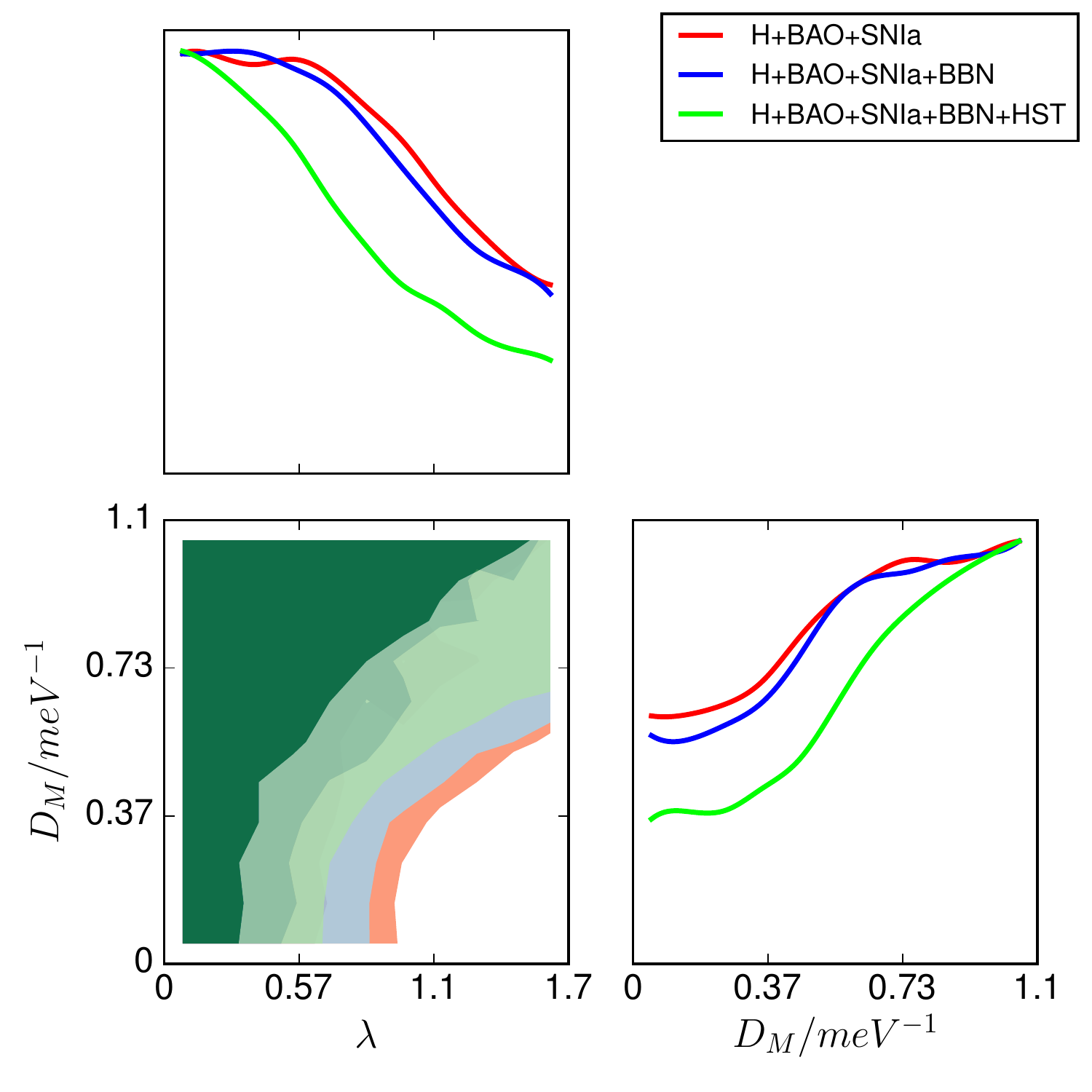}
  \caption{Confidence--level contours of model parameters for the disformally coupled DM case. We compare the $68.3\%$ (dark shaded) and $95.4\%$ (light shaded) constraints arising from $H+$SNIa+BAO observations with $H+$SNIa+BAO+BBN and $H+$SNIa+BAO+BBN+HST observations. The marginalized one--dimensional posterior distributions are also shown for comparison. } 
\label{fig:disformal_model}
\end{figure}

On the other hand, although the obtained limits on the disformal coupling might be tightened further by including higher redshift experiments, our chosen data sets indicate a preference towards a non--zero disformal coupling. In such models we find that, for a fixed potential slope $\lambda$, a weak disformal coupling $\left(D_M<\mathcal{O}\left(\text{meV}^{-1}\right)\right)$ pushes the late--time effective equation of state to $w_\phi$ or larger, i.e. $\gtrsim -1$, whereas larger disformal couplings $\left(D_M\sim\mathcal{O}\left(\text{meV}^{-1}\right)\right)$ are found to decrease the effective equation of state in the late--time Universe. Such behaviour is depicted in the top right panel of Fig. \ref{fig:couplings}. Despite the fact that different probes were used, in Ref. \cite{Xia:2013dea} they found that dynamical dark energy models with a time--dependent equation of state that cross the phantom boundary into super-acceleration are favoured by about $2\sigma$. Larger values of the scalar field potential slope $\lambda$ are allowed in comparison with the purely conformal case. We further include the HST prior and we obtain a larger disformal coupling upper limit of $D_M>0.094\;\text{meV}^{-1}$ $(95.4\%\;\text{c.l.})$. This is similar to what happened in the purely conformal case, i.e. we can tentatively say that the HST prior favours an interacting dark sector irrespective of the functional form of the dark sector coupling. The confidence--level contours and the corresponding one--dimensional posterior distributions for the different runs are shown in Fig. \ref{fig:disformal_cosmo} and Fig. \ref{fig:disformal_model}. 

%

\subsection{Mixed conformal disformal case}
\label{sec:conformal_disformal}
\begin{figure}[t]
\centering
  \includegraphics[width=1\textwidth]{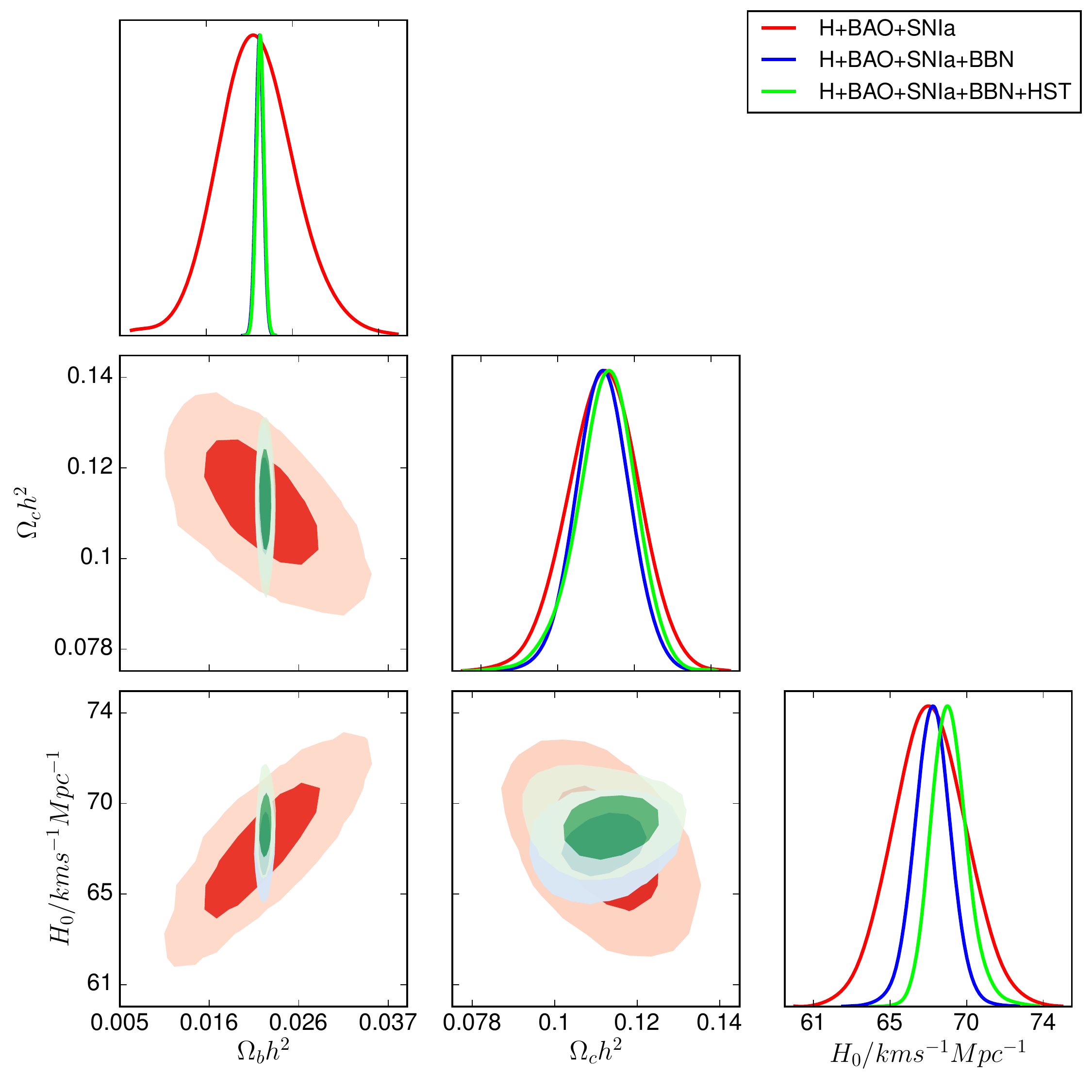}
  \caption{Confidence--level contours of the cosmological parameters in the conformally disformally coupled DM scenario. We compare the $68.3\%$ (dark shaded) and $95.4\%$ (light shaded) constraints arising from $H+$SNIa+BAO observations with $H+$SNIa+BAO+BBN and $H+$SNIa+BAO+BBN+HST observations. The marginalized one--dimensional posterior distributions are also shown for comparison.  } 
\label{fig:conformal_disformal_cosmo}
\end{figure}

\begin{figure}[t]
\centering
  \includegraphics[width=1\textwidth]{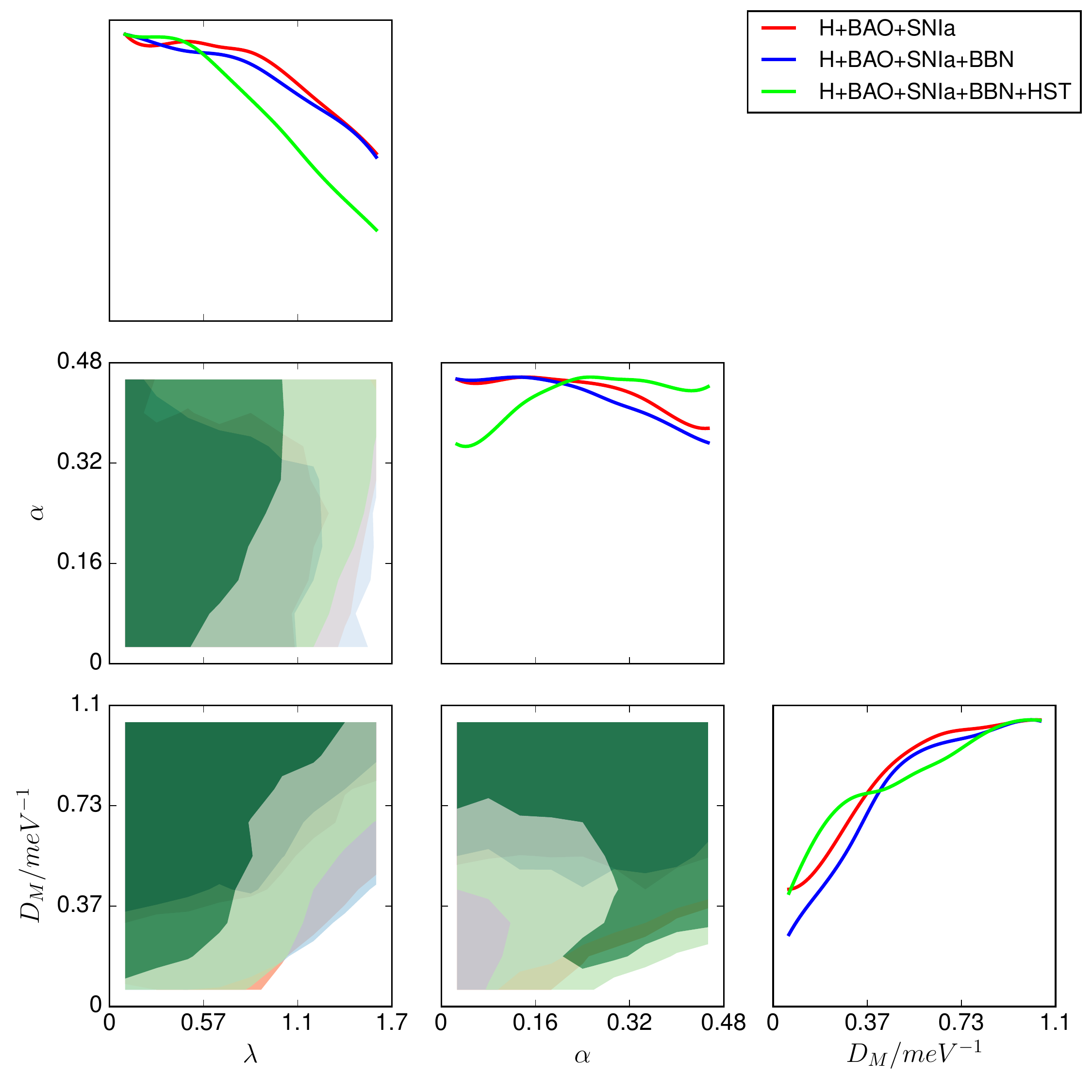}
  \caption{Confidence--level contours of the model parameters for the conformally disformally coupled DM case. We compare the $68.3\%$ (dark shaded) and $95.4\%$ (light shaded) constraints arising from $H+$SNIa+BAO observations with $H+$SNIa+BAO+BBN and $H+$SNIa+BAO+BBN+HST observations. The marginalized one--dimensional posterior distributions are also shown for comparison. } 
\label{fig:conformal_disformal_model}
\end{figure}

We now allow for both conformal and disformal couplings between dark matter and dark energy. As to be expected, the obtained constraints on parameters are weaker than those obtained in the purely conformal and the purely disformal cases presented above. We compare the results from different runs in Table \ref{table:conformal_disformal}. The obtained upper limit on the conformal coupling parameter is given by $\alpha<0.453$ $(95.4\%\;\text{c.l.})$ when using the $H$+BAO+SNIa data sets and also when including the BBN prior. When we further include the HST prior, the full range of our chosen prior is allowed, i.e. $\alpha<0.480$ $(95.4\%\;\text{c.l.})$. \emph{Hence, in the presence of an additional disformal coupling, larger conformal couplings are allowed}. In this mixed model, the lower limits on the constant disformal coupling are given by $D_M>0.102\;\text{meV}^{-1}$ $(95.4\%\;\text{c.l.})$ when using the $H$+BAO+SNIa data sets, $D_M>0.143\;\text{meV}^{-1}$ $(95.4\%\;\text{c.l.})$ when including the BBN prior, and  $D_M>0.105\;\text{meV}^{-1}$ $(95.4\%\;\text{c.l.})$ when we further add the HST prior. Again, a larger disformal coupling is preferred in comparison with the purely disformal case. The effective equation of state discussion presented in Section \ref{sec:disformal} also applies to this model. Indeed, the evolution of the effective equation of state in these models is similar to that obtained in purely disformal models. An illustration is given in Fig. \ref{fig:couplings}. The confidence--level contours and the corresponding one--dimensional posterior distributions for all the runs presented in Table \ref{table:conformal_disformal} are shown in Fig. \ref{fig:conformal_disformal_cosmo} and Fig. \ref{fig:conformal_disformal_model}. The obtained contours are much wider than those obtained in the previous models, although  high--redshift probes might shrink these contours and provide better best fits on parameters.
%
\begin{table}
\begin{center}
\begin{tabular}{ c  c  c  c } 
 \hline
\hline
 Parameter~   &  ~$H$+BAO+SNIa~ & ~$H$+BAO+SNIa+BBN~ & ~$H$+BAO+SNIa+BBN+HST~  \\ 
\hline
$\Omega_b h^2$ & $0.021^{+0.0046}_{-0.0052}$ & $0.022^{+0.0005}_{-0.0005}$ & $0.022^{+0.0005}_{-0.0005}$\\
$\Omega_c h^2$ & $0.11^{+0.011}_{-0.011}$ & $0.11^{+0.008}_{-0.008}$ & $0.11^{+0.010}_{-0.008}$\\
$H_0$ & $67.49^{+2.13}_{-2.13}$ & $67.77^{+1.10}_{-1.12}$   & $69.68^{+1.04}_{-1.16}$ \\ 
$\lambda$ & $<1.59$ & $<1.58$ & $<1.52$  \\
$\alpha$ & $<0.453$ &  $<0.453$ & $<0.480$ \\  
$D_M$    & $>0.102$ &  $>0.143$ & $>0.105$ \\
\hline
\hline
\end{tabular}
\end{center}
\caption{\label{table:conformal_disformal} For each cosmological parameter we report the best fit values and $1\sigma$ errors in the conformally disformally coupled DM scenario. For $\lambda$, $\alpha$, and $D_M\;\left(\text{meV}^{-1}\right)$, we quote the $95.4\%$ limits instead. See the top of Section \ref{sec:results} for the parameter priors.}
\end{table}
%

\section{Conclusions}
\label{sec:conclusions}
In the present work we have considered an interacting dark sector in which we allowed for two distinct forms of couplings that connect dark matter with dark energy, where the latter is responsible for the cosmological acceleration. Our current state of ignorance regarding the physics of this dark sector still allows for other interactions beyond the purely gravitational ones to exist between its elements. Various dark sector models involving various coupling functions have been extensively studied, together with their astrophysical and cosmological consequences, and it is these studies, that compare such models with state-of-the-art cosmological data, that will allow us to separate the viable candidates from the false. 

We here considered a specific coupled dark energy model in which dark energy and dark matter are allowed to couple via a conformal coupling and/or a disformal coupling. We first considered the purely conformal and the purely disformal coupling cases, and finally we also discussed the mixed scenario in which both a conformal and a disformal coupling are present. In our analysis we have only used the cosmological background evolution to constrain cosmological model parameters, namely Hubble parameter measurements, baryon acoustic oscillation distance measurements, and the Supernovae Type Ia Union2.1 compilation consisting of 580 data points. 

In the conformally coupled model, we obtained results consistent with those found in the literature, although weaker constraints were obtained as we use only the background evolution to test the models. Allowing for an additional constant disformal coupling term, we found that the constraints on the conformal coupling are relaxed. This is consistent with the observations made in Ref. \cite{Jack}, in which it was shown that the disformal term suppresses the coupling $Q$ at larger redshifts and therefore has an impact on the evolution of the effective equation of state $w_{\rm eff}$. 

We also found that, with our choice of data sets, a non-zero disformal coupling between dark matter and dark energy is preferred over the $\Lambda$CDM model. In the purely conformal coupled case, only the analysis including the HST data prefer a non-zero coupling at 1$\sigma$ confidence level. In the case of a purely disformal coupling, a non-zero coupling is preferred in all analyses, as it is the case in the conformally-disformally coupled scenario. We must now go beyond the background evolution and consider the growth of perturbations as well. Using precise measurements of CMB anisotropies and the matter power spectra of large scale structures, we certainly expect to get tighter constraints on our model parameters. We address this in future work.    

Finally, on a more speculative note, we can compare our findings above with that of Ref. \cite{Salvatelli:2014zta}, wherein Planck, SNIa, and redshift space distortion data are found to favour a late-time interaction between dark sector elements---it is shown in Ref. \cite{Jack} that the disformal coupling of the type we have just considered switches on at late times and is negligible in the past. We merely highlight this curiosity now and return to a comparison between 
the models in future work.

\acknowledgments{We would like to thank J. Lesgourgues for fruitful discussion. The work of CvdB is supported by the Lancaster- Manchester-Sheffield Consortium for Fundamental Physics under STFC Grant No. ST/L000520/1.}

\bibliography{fullbib}

\end{document}